\documentclass[5p]{elsarticle}

\usepackage{graphicx}
\usepackage{amsmath}

\newcommand{\D}{\ensuremath{d}}
\newcommand{\E}{\ensuremath{e}}
\newcommand{\imag}{\ensuremath{\mathrm{i}}}

\begin{document}

\title{Numerical approaches to time evolution of complex quantum systems} 

\author[hgw]{Holger Fehske}
\author[hgw]{Jens Schleede}
\author[hgw,rrze]{Gerald Schubert}
\author[rrze]{Gerhard Wellein}
\author[moscow]{Vladimir S. Filinov}
\author[lanl]{Alan R. Bishop}
\address[hgw]{Institut f{\"u}r Physik, Ernst-Moritz-Arndt Universit{\"a}t Greifswald, Felix-Hausdorff-Str. 6, 17487 Greifswald, Germany}
\address[rrze]{Regionales Rechenzentrum Erlangen, Friedrich-Alexander-Universit{\"a}t Erlangen-N{\"u}rnberg, Martensstr.1, 91058 Erlangen, Germany}
\address[moscow]{Joint Institute for High Temperatures, Russian Academy of Sciences, Moscow 127412, Russia}
\address[lanl]{Theory, Simulation and Computation Directorate, 
  Los Alamos National Laboratory, Los Alamos, New Mexico 87545, U.S.A.}

%

\begin{abstract}

We examine several numerical techniques for the calculation of 
the dynamics of quantum systems. 
In particular, we single out an iterative method which is based on
expanding the time evolution operator into a finite series of
Chebyshev polynomials.
The Chebyshev approach benefits from two advantages over the standard
time-integration Crank-Nicholson scheme: speedup and efficiency.
Potential competitors are semiclassical methods such as
the Wigner-Moyal or quantum tomographic approaches.
We outline the basic concepts of these techniques
and benchmark their performance against the Chebyshev approach
by monitoring the time evolution of a Gaussian wave packet in 
restricted one-dimensional (1D) geometries. Thereby the focus is on 
tunnelling processes and the motion in anharmonic potentials. 
Finally we apply the prominent Chebyshev technique to  
two highly non-trivial problems of current interest:  
(i) the injection of a particle in a disordered 2D graphene 
nanoribbon and (ii) the spatiotemporal evolution of polaron
states in finite quantum systems.
Here, depending on the disorder/electron-phonon coupling strength and
the device dimensions, we observe transmission or localisation
of the matter wave.
\end{abstract}


\maketitle

\section{Introduction}

Quantum statistical physics, such as condensed matter or plasma physics,
but also quantum chemistry, heavily depends on effective 
numerical methods for solving complex few-particle and many particle problems.
Implementing suitable theoretical concepts for their description
on modern supercomputer architectures, nowadays computational physics 
constitutes, besides experiment and theory, the third pillar of 
contemporary physics~\cite{FWS08}. 
Numerical techniques become especially important for strongly correlated systems 
where analytical approaches largely fail.
This is due to the absence of small (coupling) parameters or, more general, 
because the relevant energy scales are not well separated, both preventing 
the application of standard perturbative schemes.

Common to any numerical method in quantum physics is the requirement to
represent the states and operators describing the physical system
in the Hilbert space in a form that is suited for computations.
Then, working with a discrete basis of the Hilbert space, the computational 
challenge is the solution of an eigenvalue problem for huge (sparse) matrices.
For most physical systems the dimension of the Hilbert space is 
much too large in order to perform a full exact diagonalisation of the related 
Hamilton matrix.
Fortunately some quantities of interest depend on the properties
of the ground state or a few excited states only, and therefore may be
studied by iterative Krylov space techniques such as Lanczos 
diagonalisation~\cite{CW85}.
The quantum dynamics or long time behaviour
of correlated systems, however, require, 
in principle, the knowledge of all eigenstates. 

The theoretical investigation of quantum dynamics was triggered in 
recent years by the vast progress of the experimental techniques.
Nowadays femtosecond laser spectroscopy, for instance, allows for
a precise analysis of quantum dynamical processes with extreme
time resolution.
Direct time integration of the Schr\"odinger equation
at the cost of a full diagonalisation of the system's Hamiltonian
(including the coupling to external fields) is impractical  
in such cases because of its computational complexity.

The aim of this work is to propose a very efficient 
Chebyshev-based algorithm that allows calculating the 
dynamics of a quantum system numerically exactly, also for relatively 
long times, and therefore overcomes the above mentioned
problem at least partially. In order to demonstrate the power of 
our iterative Chebyshev expansion approach, we compare the 
accuracy and computational costs of certain model calculations 
with those emerging by the use of the more the standard Crank-Nicholson, 
Wigner-Moyal and quantum tomography methods. 
We start by presenting the basic ideas of the iterative 
(Chebyshev, Crank-Nicholson; Sect.~\ref{sect:directmethods}) and 
semiclassical (Wigner-Moyal, tomographic; 
Sect.~\ref{sect:semiclassicalmethods}) techniques.
Afterwards, in Sect.~\ref{sect:systems}, we consider three different 
problems of increasing complexity: (i) the motion of a Gaussian wave 
packet in a 1D geometry (Sec.~\ref{sect:doublewells}), 
(ii) the evolution of a 
particle in a disordered 2D graphene nanoribbon (Sec.~\ref{sect:grapheneribbon}), 
and (iii) the spatiotemporal evolution of polaron states in 
finite quantum systems (Sec.~\ref{sect:polaron}). 
Our conclusions will be presented in Sect.~\ref{sect:conclusion}.

\section{Time evolution of quantum systems}
\label{sect:methods}

The time evolution of a quantum state $|\psi\rangle$ is 
described by the Schr{\"o}dinger equation
\begin{equation}
  \imag \hbar \frac{\partial}{\partial t}|\psi(t)\rangle = H |\psi(t)\rangle\,.
\end{equation}
If the Hamilton operator $H$ does not explicitly depend on time $t$ 
we can formally integrate this equation and express 
the dynamics of a given quantum state $|\psi(t_0)\rangle$
in terms of the time evolution operator $U(t,t_0)$ as 
$|\psi(t)\rangle = U(t,t_0)|\psi(t_0)\rangle$, where $U(t,t_0) =
\E^{ -\imag H (t-t_0)/\hbar}$.
Exploiting that $U(t,t_0)$ is diagonal in the eigenbasis of the 
Hamiltonian, we can directly determine the dynamics  
of the quantum system.

\subsection{Direct method}
\label{sect:directmethods}

For systems with moderate Hilbert space dimensions, a full 
diagonalisation of the Hamiltonian permits expression of the quantum 
dynamics of an initial state $|\psi(t_0)\rangle$ as
\begin{equation}
  |\psi(t)\rangle = \sum\limits_{n=1}^{N} e^{-\imag E_n (t-t_0)/\hbar} 
  |n\rangle\langle n| \psi(t_0)\rangle\,.
\end{equation}
Here $|n\rangle$ are the (time independent) eigenstates of the system and $E_n$
the corresponding eigenenergies.
In this way, the decomposition of an initial state into a linear combination of 
eigenstates allows for an exact calculation of the quantum state at 
arbitrary times.
As soon as the physical description of the system requires a larger
Hilbert space dimension, however, 
this direct calculation is no longer feasible and
we have to resort to alternative approaches.

\subsection{Iterative methods}
\label{sect:iterative}
{\sl Crank-Nicholson scheme.} 
One of the standard methods to calculate the quantum time evolution
iteratively is the Crank-Nicholson algorithm~\cite{PFTV86}.
Dividing $[t_0,t]$ into $S$ subintervals $\delta t$, the quantum state
evolves for each iterative time step $\delta t$ from $t_s$ to 
$t_{s+1}=t_s+\delta t$ according to
\begin{equation}
  \Big(1+\frac{1}{2}\imag H \delta t/\hbar \Big)  |\psi(t_{s+1})\rangle = 
  \Big(1-\frac{1}{2}\imag H \delta t/\hbar \Big)  |\psi(t_s)\rangle\,.
  \label{eq:CN}
\end{equation}
There are two limitations to this scheme. 
First, in addition to the matrix vector multiplication (MVM) on the right
hand side of (\ref{eq:CN}), each iteration requires the solution of a linear 
system of equations to obtain $|\psi(t_{s+1})\rangle$.
Despite the availability of many powerful methods for the solution of large 
(sparse) linear equation systems, this task remains the most time consuming 
part of the algorithm. 
Using iterative methods for the solution of the linear equation system, the
attainable Hilbert space dimensions increase substantially 
as compared to direct methods.
Second, the Crank-Nicholson algorithm is accurate only to order $(\delta t)^2$,
which severely restricts the maximum usable iterative time step.

{\sl Chebyshev scheme.} 
Both limitations can be overcome by an approach where we 
expand the time evolution operator $U(t,t_0) = U(t-t_0)=U(\Delta t)$
into a finite series of first-kind Chebyshev polynomials of order $k$:  
$T_k(x)=\cos(k\, \mathrm{arccos}(x))$.
We then obtain~\cite{CG99,TK84,WF08}
\begin{equation}
  U(\Delta t) =  \E^{-\imag b \Delta t/\hbar} 
  \Big[ c_0(a\Delta t/\hbar) + 2\sum\limits_{k=1}^{M} c_k(a\Delta t/\hbar)
    T_k(\tilde{H}) \Big].
\label{eq:U_1}
\end{equation}
Prior to the expansion the Hamiltonian has to be shifted and rescaled such that
the spectrum of $\tilde{H} = (H-b)/a$ is within the definition interval of the
Chebyshev polynomials, $[-1,1]$.
The parameters $a$ and $b$ are calculated from the extreme eigenvalues of $H$ as
$b=\frac{1}{2}(E_{\mathrm{max}}+E_{\mathrm{min}})$ and 
$a=\frac{1}{2}(E_{\mathrm{max}}-E_{\mathrm{min}}+\epsilon)$.
Here we introduced $\epsilon=\alpha(E_{\mathrm{max}}-E_{\mathrm{min}})$ to ensure 
the rescaled spectrum $|\tilde{E}| \le 1/(1+\alpha)$ lies 
well inside $[-1,1]$. In practice, we use $\alpha=0.01$. 
Note that the Chebyshev expansion also applies 
to systems with unbounded spectra.
In those cases we truncate the infinite Hilbert space to a finite 
dimension by restricting the model on a discrete space grid or using an energy
cutoff.
In this way we ensure the finiteness of the extreme eigenvalues.

In~(\ref{eq:U_1}), the expansion coefficients $c_k$ are given by 
\begin{equation}
  c_k(a\Delta t/\hbar) = \int\limits_{-1}^1 
  \frac{T_k(x)e^{-\imag x a \Delta t/\hbar }}{\pi \sqrt{1-x^2}}\D x =
  (-\imag)^k J_k(a \Delta t/\hbar)
\end{equation}
($J_k$ denotes the $k$-th order Bessel function of the first kind). 

To calculate the evolution of a state $|\psi(t_0)\rangle$ from one 
time grid point to the adjacent one,
$|\psi(t)\rangle = U(\Delta t)|\psi(t_0)\rangle$, we
have to accumulate the $c_k$-weighted vectors
$|v_k\rangle = T_k(\tilde{H})|\psi(t_0)\rangle$.
Since the coefficients $c_k(a\Delta t/\hbar)$ depend on the time step but not
on time explicitly, we need to calculate them only once.
The vectors $|v_k\rangle$ can be computed iteratively exploiting
the recurrence relation of the Chebyshev polynomials,
\begin{equation}
    |v_{k+1}\rangle = 2\tilde{H} |v_k\rangle - |v_{k-1}\rangle\;,
\end{equation}
with $|v_1\rangle = \tilde{H} |v_0\rangle$ and $|v_0\rangle = |\psi(t_0)\rangle$.
Evolving the wave function from one time step to the next
requires $M$ MVMs  of a given complex vector with 
the (sparse) Hamilton matrix of dimension $N$ 
and the summation of the resulting vectors 
after an appropriate rescaling.
The Chebyshev expansion may also be applied to systems with
time dependent Hamiltonians, but there the time variation 
$H(t)$ determines the maximum $\Delta t$ by which
the system may be propagated in a single time step.
For time independent $H$, in principle, arbitrary large time steps are
possible at the expense of increasing $M$.
We may choose $M$ such that for $k>M$ the modulus of all expansion coefficients 
$|c_k(a\Delta t/\hbar)|\sim J_k(a\Delta t/\hbar)$ is smaller than a desired
accuracy cutoff.
This is facilitated by the fast asymptotic decay of the Bessel functions, 
\begin{equation}
  J_k(a\Delta t/\hbar)
  \sim \frac{1}{\sqrt{2\pi k}} \left( \frac{\E a\Delta t}{2\hbar k} \right)^k 
  \quad {\mathrm {for}}\quad k\to \infty\;.
  \label{eq:dec_Bessel}
\end{equation}
Thus, for large $M$, the Chebyshev expansion can be considered as 
quasi-exact, and permits a considerably larger time step than
e.g. the Crank-Nicholson scheme.
Besides the high accuracy of the method, the linear scaling of 
computation time with both time step and Hilbert space dimension are 
promising in view of potential applications to more complex systems.
In our cases almost all computation time is spent in sparse MVMs, 
which can be efficiently parallelised, allowing for a good speedup on 
parallel computers.

\subsection{Semiclassical methods}
\label{sect:semiclassicalmethods}
During the last decades, a variety of semiclassical methods have been
tailored in order to incorporate certain quantum effects at least partially 
into classical many-particle simulations.
Based on the real time (Feynman-) path integral formulation of 
quantum mechanics, where action integrals take the
center stage, they allow propagation of the (complex) wave function of a
quantum system in time.
Within the numerical evaluation of the  integrals occurring by Monte
Carlo (MC) techniques~\cite{Fi86}, the oscillatory complex 
valued integrand causes a dynamical sign problem which spoils 
the efficiency of the MC integration.

{\sl Wigner-Moyal approach.} 
Since a quantum system can be described equivalently in terms of 
real valued quantum phase space distribution functions~\cite{Le95},
e.g., the Wigner function, the dynamical sign problem may
be alleviated~\cite{Fi96c,FMK95}.~\footnote{Note that the Wigner function is just a convenient mathematical tool 
for the description of quantum systems and cannot be considered as a 
joint probability due to its possibly negative values, and conflict with  
Heisenberg uncertainty relation.}
Starting from the von Neumann equation we may derive an evolution 
equation for the Wigner function $W(q,p,t)$~\cite{Fi96c}
\begin{equation}
\frac{\partial W}{\partial t}+ \frac{p}{m} \frac{\partial W}{\partial q} 
+ F(q) \frac{\partial W}{\partial p} = \int\limits_{-\infty }^\infty 
 \D s\;W(q, p-s, t) \omega(s,q)\,,
\label{eq:EOM_Wigner}
\end{equation}
where $F(q)=-\frac{dV(q)}{dq}$ is the classical force, and 
\begin{equation}
  \omega(s,q) = \frac{2}{\pi\hbar^2}
  \int \D q'\; V(q-q') \sin \Big( \frac{2 sq'}{\hbar}\Big) + 
F(q) \frac{d\delta(s)}{ds}\,.
  \label{eq:omega}
\end{equation}
In the classical limit, the right hand side of (\ref{eq:EOM_Wigner}) vanishes, leaving
us with the Liouville equation for the phase space density.
Then the  dynamics can be expressed in terms of the classical 
propagator
\begin{equation}
\begin{split}
  \Pi^W(q,p,t;q_0,p_0,t_0) = & \delta [q-\bar{q}(t;p_0,q_0,t_0)]\\
  & \times\delta [p-\bar{p}(t;p_0,q_0,t_0)] 
  \,,
\end{split}
\end{equation}
where $\bar{p}$ and $\bar{q}$ are the momentum and coordinate 
of a trajectory that evolves according to the Hamilton 
equations of motion with initial
conditions $\bar{p}(t_0)=p_0$ and $\bar{q}(t_0)=q_0$.
Using $\Pi^W$, we may rewrite (\ref{eq:EOM_Wigner}) in form of an 
integral equation~\cite{Fi96c,FMK95},
\begin{equation}
\begin{split}
W(q,p,t) = &\int  \D p_0\,  \D q_0 \;\Pi^W(q,p,t;q_0,p_0,t_0)
W_0(q_0,p_0,t_0)  \\&
 +  \int\limits_{t_0}^t d\tau \int \D p_{\tau} \D q_{\tau}\;
\Pi^W (q,p,t; q_{\tau},p_{\tau},\tau)  \\&\quad\times 
\int\limits_{-\infty }^\infty \D s\; W(q_{\tau}, p_{\tau}-s,\tau)
 \omega (s,q_{\tau})  \label{eq:s6}\,,
\end{split}
\end{equation}
and solve it by iteration. 
Here, we consider only the lowest order, which means that the
second integral is neglected completely.
That is, we propagate classical trajectories $(\bar{q},\bar{p})$ in time,
after sampling their initial conditions $p_0$ and $q_0$ from the initial
Wigner function $W_0(q_0,p_0,t_0)$ at time $t_0$ by a MC procedure.
Their superposition at the next time grid point gives $W(q,p,t)$.
The importance of higher order terms in the iteration series was 
investigated in Refs.~\cite{Fi96c,FMK95}.

{\sl Quantum tomographic approach.} As  the dynamical sign problem is 
still present for the Wigner function, some years ago 
the description of quantum states in terms 
of a strictly positive function, the so called quantum tomogram,
has been proposed~\cite{MMT9697}.
Such a description seems promising in view of an effective MC sampling
of the trajectories during the propagation. 
The quantum tomogram~\cite{AL03,ALM04}, 
\begin{equation}
\label{eq:tom}
  \tilde{w}(X,\mu,\nu,t) = \int \frac{\D k\, \D q\, \D p }{2\pi}
    W(q,p,t) \E^{-\imag k(X - \mu q - \nu p)}\,,
\end{equation}
relates to the Wigner function by a class of Radon transforms~\cite{De83}
which are characterised by $\mu$ and $\nu$.
Each tomogram contains a density information, and
tuning $(\mu,\nu)$ appropriately, we may continuously 
switch between coordinate and momentum representation.
Also for the quantum tomogram, an evolution equation can be derived
from the von Neumann equation 
\begin{equation}
  \begin{split}
  \frac{\partial \tilde{w}}{\partial t} -
  \frac{\mu}{m}\frac{\partial \tilde{w}}{\partial \nu} -
  \frac{\imag}{\hbar}  \left[V \left(
      - \frac{\partial }{\partial \mu} \frac{1}{\partial / \partial X}
      - \frac{ \imag\hbar\nu}{2} \frac{\partial }{\partial X} \right)\right.\\
  \left. -
    V\left( - \frac{\partial }{\partial \mu} \frac{1}{\partial /
        \partial X} + \frac{ \imag\hbar\nu}{2} \frac{
        \partial }{\partial X} \right) \right ] \tilde{w} = 0\,.
  \label{eq:EOM_TOM}
  \end{split}
\end{equation}

For harmonic potentials, Eq.~(\ref{eq:EOM_TOM}) can be reformulated as
a continuity equation and solved by trajectory methods.
For potentials of arbitrary shape, the complicated structure 
prevents a direct evaluation of (\ref{eq:EOM_TOM}) in order to get 
$\tilde{w}(X,\mu,\nu,t)$.
A possible way out is a local expansion of the potential up to 
second order~\cite{SFMSF08p}.
The identification with the continuity equation then holds locally
for each coordinate about which the potential is expanded.
Since the slope and local curvature of the potential enter the 
propagator for the trajectories, the efficiency of the
tomogram-reconstruction depends crucially on the choice of
the potential sampling positions.
An intuitive, albeit not unique, choice for those positions are
the coordinates of classically evolving particles similar to 
those in the Wigner approach.
Advantageously, it is not necessary to propagate the whole set of tomograms
if one is interested in the tomogram in one reference frame $(\mu,\nu)$
only.
Instead, it is sufficient to find those trajectories which 
end up in a given $(X(t),\mu(t),\nu(t))$, or equivalently to propagate the
desired $(X,\mu,\nu)$--trajectories backward in time: one needs 
to be aware of the non-uniqueness of the propagator 
due to the various possible choices of the potential sampling points. 


\section{Topical applications in physics}
\label{sect:systems}
We now apply the numerical techniques presented in the preceding
section to selected physical systems and situations. 
As a first step, we calibrate the different approaches by 
studying a simple double well toy model.
Detecting the limitations and prospects of the various methods
seems to be necessary before applying them to the more complicated 
problems of current interest.
Let us point out that the considered implementations of
the semiclassical approaches provide an approximate 
description of quantum mechanics only, i.e. they 
will clearly not reproduce all the quantum effects.
For the finite graphene nanoribbons studied in the second example 
it is barely possible to determine the full quantum dynamics 
by means of direct diagonalisation techniques or applying 
the Crank-Nicholson scheme.
This gives us the opportunity to benchmark their performance in
comparison to the Chebyshev expansion for a system for which 
the Hamilton matrix is not tridiagonal.  
The last example has been chosen to demonstrate the applicability of 
the Chebyshev approach to a true many-particle problem, the tunnelling of
a polaron. 
There the Hilbert space dimension is so large that neither the direct 
diagonalisation method nor the Crank-Nicholson can be applied.
%
\subsection{Double well potential}
\label{sect:doublewells}
\begin{figure}
  \centering
  \includegraphics[width=0.9\linewidth,clip]{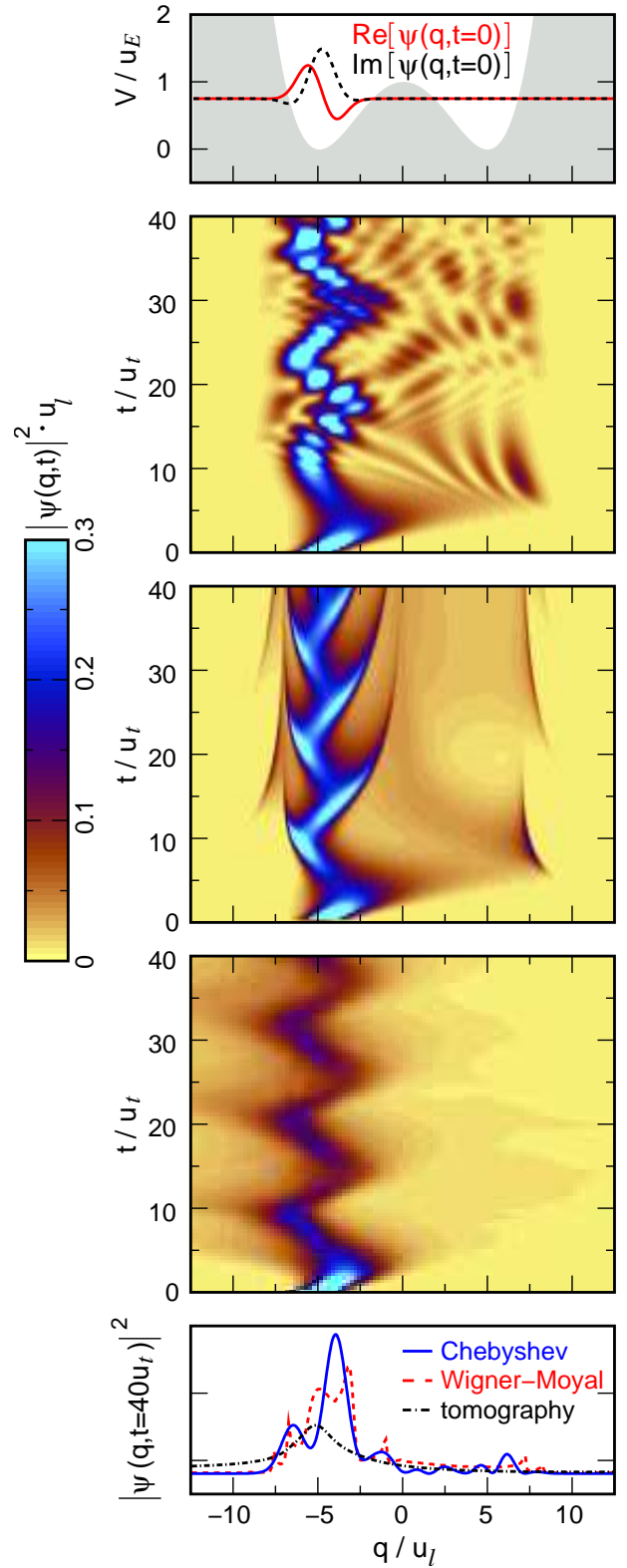}
   \caption{Time evolution of a Gaussian wave packet in a double well
   potential (top panel). Results displayed are obtained using the Chebyshev, 
   the first order Wigner-Moyal and the tomographic approach (from 
   top to bottom), respectively.}
   \label{fig:doublewell}
\end{figure}

As a basic test case we consider the motion of a Gaussian wave packet in the 
1D double well potential
\begin{equation}
   V(q) = V_0 + \frac{1}{2}m\omega_0^2 (-q^2 + a q^4)\,,
\end{equation}
sketched in the top panel of Fig.~\ref{fig:doublewell}.
We use $m=u_m$, $\;\omega_0 u_t=0.4$, $\;a u_\ell^2 =0.02$ and 
$V_0=u_E=u_mu_\ell^2/u_t^2$, where $u_m$, $u_t$, $u_\ell$ are 
the reference units for mass, time and length, respectively. 
The initial Gaussian of width $\sigma$ is centred at $q_0$, 
with centre-of-mass momentum $p_0$,
\begin{equation}
  \psi(q,t_0) =  \frac{1}{(2\pi\sigma^2)^{1/4}} 
  \exp\left\{-\frac{1}{4\sigma^2} (q-q_0)^2 + \frac{\mathrm i}{\hbar} p_0 q\right\}\,,
\label{eq:psi0}
\end{equation}
where we choose  $p_0 = u_m u_\ell/u_t$,  $q_0 = -5 u_\ell$, and 
$\sigma=u_\ell/\sqrt{2}$.
The top panel of Fig.~\ref{fig:doublewell} gives also the real and 
imaginary part of the wave function, where the baseline indicates 
the energy of the initial state in relation to $V(q)$.

Discretising the potential $V(q)$ on an equally spaced grid of $N=2048$ 
$q$--points, we fully diagonalise the Hamiltonian to  
get the exact dynamics as reference.
If we choose the iterative time step accordingly, 
the real and imaginary part of the wave function at $t=40 u_t$
is reproduced by both iterative methods with an absolute error less 
than $10^{-9}$.
For this the maximum allowable time step for the Crank-Nicholson scheme is 
$\delta t=4\times 10^{-6}u_t$. 
For the Chebyshev approach, the accuracy is even better 
than $2\times10^{-11}$.
Here the required time step is related to the order
of the Chebyshev expansion, i.e. the number of moments $M$.
For $M=1500$ we may propagate the wave function by $\Delta t=0.4u_t$.
At the expense of calculating a larger number of moments  $M$, 
larger times steps may be chosen without loss of accuracy. 
This will be demonstrated in the next  section.
The time evolution of the modulus squared of the wave function is
shown in the second panel of Fig.~\ref{fig:doublewell}.
While the major part of the wave packet stays in the left well, a sizeable 
fraction also penetrates the barrier.
The rich structure in the density pattern reflects the (well-known) 
presence of strong interference effects.

Restricting the Wigner-Moyal scheme to first order, the corresponding 
data  reproduces in essence the overall dynamics, but 
the fine interference patterns are not resolved within this approximation
(see third panel of Fig.~\ref{fig:doublewell}).
There are two major parameters which influence the computation time for this
approach: (a) the number of propagated trajectories and (b) the 
time step necessary for their classical propagation.
For the results presented, we have used $5\times10^{5}$ trajectories 
and a propagation time step of $0.04u_t$.
The agreement between the exact solution and the tomographic result 
(see also the bottom panel of Fig.~\ref{fig:doublewell}) is even 
worse as the tunnelling to the right hand
side of the barrier is missed almost completely.
Furthermore, the probability density for large negative values is 
overestimated, i.e. the effect of the steep anharmonic confinement 
potential is not accounted for correctly.
In Tab.~\ref{tab:doublewell} we summarise the run times $T_{\mathrm{run}}$
on a single Xeon 5160 processor required by the different methods in order
to follow the time evolution of the system up to $t=40 t_0$.

This very basic example already shows that a straightforward 
use of the Wigner-Moyal and tomographic approaches 
only partially accounts for quantum effects.
While, in principle, both methods are equivalent with respect to 
the solution of the time dependent Schr{\"o}dinger equation, an 
efficient implementation is lacking.
For the Wigner-Moyal formalism there are two prospects. 
If the accuracy of the first order approximation is satisfactory, 
i.e. the neglect of the fine interference patterns is tolerable,
this method provides an acceptable performance and might show its true 
virtue for many-particle systems.
If one has to include higher terms of the iteration series, however, 
e.g. because subtle quantum effects are important, the method
is not competitive anymore because (i) the computational requirements
increase drastically and (ii)  numerical fluctuations amplify strongly 
during the calculation~\cite{SFMSF08p}.
The practical applicability of the tomographic approach to arbitrarily 
shaped potentials is also questionable, mainly because there
is no simple way to construct suitable sampling functions for 
the coordinate sampling in the potential evaluation.
Extensions beyond harmonic potentials will suffer 
crucially from this limitation.
Apart from the poor accuracy of the results, also the computational 
requirements were significant higher than for the other methods  
(although we used only $800$ trajectories in this calculation).

 \begin{table}
   \centering
   \begin{tabular}{|c||c|c|c|c|c|}\hline
     & ED & CN & C & WM & T\\\hline
     $T_{\mathrm{run}}[s]$&13.8 & 1871  & 4.8 & 13.6 & 579 \\\hline
   \end{tabular}
   \caption{Time evolution of a wave packet in the double well 
     potential up to $t=40t_0$.
     Data gives the run times by exact diagonalisation (ED), 
     Crank-Nicholson (CN), Chebyshev (C), Wigner-Moyal (WM) 
     and tomographic (T) methods.}
   \label{tab:doublewell}
 \end{table}

\subsection{Disordered graphene nanoribbons}
\label{sect:grapheneribbon}
Recently much interest has been devoted to investigate how 
disorder influences the transport properties of graphene~\cite{PGLPC06,XX07}.
It is known that the presence of arbitrarily weak disorder 
leads to Anderson localisation of the single particle wave function 
on infinite 2D square lattices~\cite{AALR79}.
In weakly disordered 2D systems the localisation 
lengths are huge, however, and may easily become comparable 
to the system sizes that are technologically relevant e.g. for 
graphene nanoribbons. 
In those cases, we expect a conducting behaviour of the device
despite the presence of disorder causing localisation on larger 
length scales.  

\begin{figure}[t]
  \centering
  \includegraphics[width=\linewidth,clip]{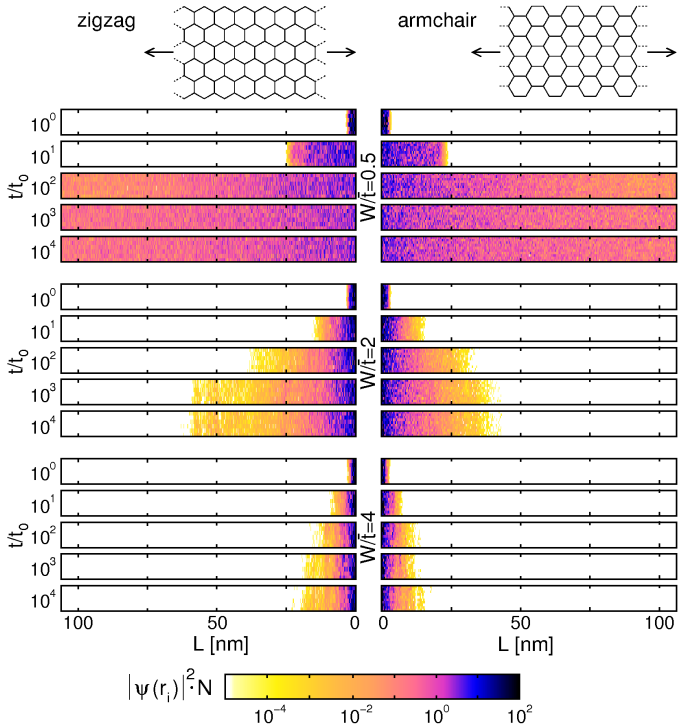}
  \caption{Time evolution of a localised state on 
    an armchair graphene nanoribbon for different values of disorder. 
    The left/right asymmetry of the state is due to the particular
    disorder realisation and initial state.
    \label{fig:graph}}
\end{figure}

To investigate the influence of Anderson disorder for finite 
graphene nanoribbons, we consider a tight-binding model on a honeycomb lattice,
\begin{equation}
  \label{eq:model}
  {H} =  - \bar{t} \sum\limits_{\langle ij \rangle} 
  \bigl({c}_i^{\dag} {c}_j^{} + \text{H.c.}\bigr)
  + \sum\limits_{j=1}^{N} \epsilon_j {c}_j^{\dag} {c}_j^{} \;.
\end{equation}
Here the operators ${c}_j^{\dag}$ (${c}_j^{}$) create (annihilate) an 
electron in a Wannier state centred at site $j$.
The on-site potentials $\epsilon_j$ are random variables in the interval 
$[-W/2,W/2]$, where $W$ is a measure for the disorder strength.
The electron transfer between nearest-neighbour lattice sites 
$\langle i j\rangle$ is described by the transfer integral $\bar{t}$.
Tailoring stripes (ribbons) out of the infinite honeycomb-lattice, we have
to distinguish two cases with respect to the boundary conditions. 
Depending on the orientation of the stripe we get boundaries of either  
zigzag or armchair type. 
Since in experimental probes armchair edges are more common, we
will focus on those in the following.

Starting from a wave packet which is localised on one site in the centre of 
each ribbon, we evolve the quantum state using the Chebyshev approach 
described in Sect.~\ref{sect:iterative}.
%
We consider devices of $1.11\times 212.8$ nm$^2$ with $10\times1000$ atoms.
The main panel of Fig.~\ref{fig:graph} displays the time evolution
of the wave packet for one realization of disorder for 
several values of disorder strength (time is measured in units of the 
inverse hopping element, $t_0=1/\bar{t}$).

The following aspects of the wave function dynamics should be noted:
(i) The initially localised wave function spreads with time and reaches its
maximum extension at about $t\simeq 10^3 t_0$, 
independent of the disorder strength.
Also for much longer times ($t\sim10^4t_0$) this extension does not change
significantly anymore. 
(ii) The disorder strength strongly influences the spatial region over 
which the state spreads, i.e. the localisation length.
While we see clear evidence for localisation at large disorder ($W=4\bar{t}$), 
the localisation length for weak disorder ($W=0.5\bar{t}$) is 
markedly larger than the system size, leading to an evenly spread state
on the ribbon.
Note that the occurrence of an extended (conducting) state in the 
latter case is only due to the finite system size. 
For longer ribbons a disorder of $W=0.5\bar{t}$ is sufficient 
to localise the wave function as well.
%

The disordered nanoribbon setup gives us a good opportunity to benchmark the
Chebyshev approach against the Crank-Nicholson scheme.
Since the Hamiltonian is no longer tridiagonal the solution of the linear 
equation system cannot be done by the Thomas algorithm.
Instead, we use the standard solver for double-complex 
linear equation systems from LAPACK, ZGESV.
Table~\ref{tab:gra} summarises the number of moments required
to get agreement between the Chebyshev and the exact results 
for different time steps $\Delta t$.
The computation times for calculating the quantum state at time $t=10^4 t_0$
using the various $\Delta t$ is also given. 
For comparison, we also give run time $T_{\mathrm{run}}$ for the
exact diagonalisation.
As one iteration of the Crank-Nicholson scheme using ZGESV takes 4 minutes,
the given $T_{\mathrm{run}}$ is only an estimate.
%

 \begin{table}
   \centering
   \begin{tabular}{|c||c|c|c||c||c|}\hline
     & \multicolumn{3}{c||}{C}& 
     CN & ED\\\hline
      $\Delta t/t_0$ & $1$ & $10$  & $100$ & $10^{-3}$ & $\infty$ \\\hline
      $M$& 104 & 264 & 1210 & $-$ & $-$ \\\hline
     $T_{\mathrm{run}}[s]$&  378 & 97 & 45 & $>10^{7}$ &  1278 \\\hline

    
   \end{tabular}
   \caption{Numeber of Chebyshev moments (M) and overall runtime
     ($\Delta t/t_0$) required for the calculation of the time evolution up to $t=10^ 4t_0$
     on a single Intel Xeon 5160 core.\label{tab:gra}}

 \end{table}

\subsection{Spatiotemporal evolution of polaron states
in finite quantum structures}
\label{sect:polaron}
The cycle of quasiparticle formation is fundamental to many fields
of physics. In condensed matter, e.g., the coupling between 
a charge carrier and the lattice degrees of freedom may create  
a new quasiparticle, an electron dressed by an phonon 
cloud. This composite entity is called polaron. From the basic
electron-phonon (EP) interaction processes, the absorption/emission
of a phonon with a simultaneous change of the electron state, it is
clear that the motion of even a single electron in a deformable
lattice constitutes a complex many-body problem, in that phonons 
are excited at various positions, with highly non-trivial 
dynamics~\cite{KT07}.  
Polaron transport through finite quantum systems becomes increasingly 
important for nanotechnology applications.  

The microscopic structure of polarons is very rich. 
Focusing on polaron formation in systems with short-range non-polar
EP interaction and site-dependent potentials, we consider a
generalised Holstein Hamiltonian~\cite{FWLB08}
\begin{eqnarray}
\label{eq:ghm}
H &=& \sum_{i}\Delta_i n_{i} - \bar{t}\sum_{i}
(c^{\dagger}_{i}c^{}_{i+1}
+\mbox{H.c.})\nonumber\\
&&-\sum_{i,\sigma}\bar{g}_i\omega_0
(b_i^{\dagger}+b_i^{})n_{i\sigma}+\omega_0\sum_{i} b_i^{\dagger} b_i^{}\,,
\end{eqnarray}
where $c^{\dagger}_i$ ($b_i^{\dagger}$) creates an electron (phonon) at 
site $i$ of a 1D lattice. Parameters are the electron transfer integral
$\bar{t}$, the EP coupling strength 
$\bar{g}_i=[(\varepsilon_{p}+\varepsilon_{p,i})/ \omega_0]^{1/2}$,
and the phonon frequency $\omega_0$. The potentials 
$\Delta_i$ can describe a tunnel barrier, disorder, or a voltage basis. 

\begin{figure}[h]
  \centering
  \includegraphics[width=\linewidth,clip]{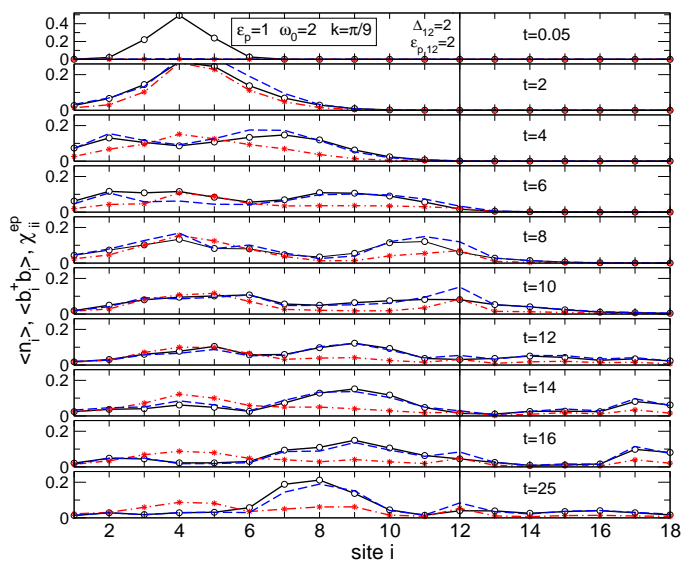}
   \caption{Quantum dynamics of polaron formation and polaron
tunnelling through a potential barrier. A Gaussian wave packet centred
around site 4 begins evolving at time $t=0$ with momentum $k=\pi/9$. Moving
to the right, a polaron is formed that hits the barrier located at site
12 at about $t\simeq 10$. There complicated reflection and
transmission processes of the composite quasiparticle made up of an
electron and phonons take place. Displayed is the time-evolution of
the local particle densities $\langle n_i\rangle$ (solid lines, open
circles), phonon numbers $\langle b_i^\dagger b_i^{}\rangle$
(dot-dashed lines, stars) and EP correlations $\chi^{ep}_{ii}=\langle
n_i^{} (b_i + b_i^+)\rangle$ (dashed lines). Open boundary conditions
were used at sites 1 and 18.  In the numerics we account for all
states with up to $N_{\text{ph}} \le 11$ phonons and in the ground state the
weight of all basis states containing exactly $N_{\text{ph}}=11$ is less than
$10^{-11}$.}
   \label{fig:polarondynamics}
\end{figure}

How does a bare electron time evolve to become a polaron
quasiparticle?  To what extent can a polaron tunnel through a quantum
barrier? Having the iterative Chebyshev-based time evolution algorithm
explained in Sec. 2.2 at hand, we can address these questions in the
framework of model~(\ref{eq:ghm}). Let us emphasise that our numerical
approach, acting in the tensorial product Hilbert space of electron
and phonons, takes into account the full dynamics of both quantum
objects.  Since the Hilbert space associated to the phonons is
infinite, we applied a controlled truncation
procedure retaining only basis states with at most
$N_{\text{ph}}$ phonons~\cite{JF07,WRF96}. However the 
truncated Hilbert space dimension ($D^{tot}$)
is still very large even for small lattices and the dimension of the
corresponding sparse matrix problem does limit the physical parameter
region attainable. Thus, we use a memory saving implementation of the
sparse MVM where the non-zero matrix elements
are not stored but recomputed in each sparse MVM step, limiting the
overall memory consumption of our implementation to five vectors of
size $D^{tot}$. In this context we can access a massively parallel
sparse MVM code which has proven to be sufficient to compute the
ground state of the model~(\ref{eq:ghm}) up 
to $D^{tot}=3.5\times10^{11}$ very efficiently on more than 5000
processor cores~\cite{FAW09}. For the single polaron dynamics presented
here, the matrix dimension is $D^{tot}=6.2\times10^8$ and we run the
Chebyshev approach on 18 processors of a SGI Altix4700 compute server
accessing a total of approximately 60 GBytes of main memory and
consuming less than 500 CPU-hrs to compute the result presented in
Fig.~\ref{fig:polarondynamics}.

Figure~\ref{fig:polarondynamics} shows snapshots of polaron formation
and polaron propagation at intermediate EP coupling $\varepsilon_p =1$ 
in the non-adiabatic regime $\omega_0=2$ (the time and all energies were  
measured in units of $\bar{t}^{-1}$ and $\bar{t}$, respectively).
At $t=0$ a bare electron wave packet 
is injected at site 4 and launched to the right. Shortly after, the electron 
is not yet dressed and moves nearly as fast as a free particle.
But then the electron emits (creates) phonons in the vicinity of 
the electron's starting point, in order to reduce its energy
to near the bottom of the band, and then forms a polaron 
(see the panel at $t=6$). One of the most important properties 
of the polaron is an increased inertial mass, for the reason that 
some phonons have to travel with the particle (as indicated 
by the enhanced on-site EP correlations). At the same time we 
observe a ``backscattered'' current~\cite{KT07},
evolving to a left moving polaron. When the right-moving polaron reaches 
the wall at site 12 it will be partly reflected. 
More importantly, the additional local EP interaction $\varepsilon_{p,12}$ 
renormalises the on-site adiabatic potential at site 12, i.e. leads 
to a local polaron level shift that softens the barrier $\Delta_{12}=2$. 
As a result a vibration-mediated tunnelling of the particle takes place, 
whereby some phonons are stripped at the barrier and are  
recollected  by the transferred particle afterwards (cf. the snapshots
from $t=10$ to 14). Finally, the particle is reflected at the boundary
and moves to the left passing the barrier again. Note that during the 
whole run time uncorrelated phonon excitations remain in the system,
especially near the injection point. 
In our opinion this example impressively demonstrates that our 
approach can be used to monitor the complicated multiple 
time-scale dynamics of quasiparticle 
transport in finite quantum structures.


\section{Conclusion}
\label{sect:conclusion}

To summarise, in this work we compared various numerical approaches to 
the dynamics of complex quantum systems: 
expansion into eigenstates, iterative Crank-Nicholson and Chebyshev 
schemes, as well as semiclassical Wigner-Moyal and quantum tomography methods.  
The different methods have been applied to several physical systems
and problems, ranging from motion in a simple double-well toy model 
to questions of current interest such as electron transport in disordered 
graphene nanoribbons or polaron motion in a finite quantum structure.

The Wigner-Moyal approach, evaluated in first order of the iteration series,
essentially reproduces the quantum dynamics. Nevertheless important 
quantum interference effects, appearing in the exact solution, are missed.
The successful application of the quantum tomogram to the time evolution 
of quantum systems crucially depends on a suitable sampling algorithm for the 
coordinates at which the potential is evaluated.
If those are sampled according to trajectories of classically propagated 
particles, the result for the quantum dynamics is rather poor and fails to 
describe tunnelling and anharmonicity effects correctly.
While the moderate numerical costs of the first order 
Wigner-Moyal approach seem appealing for a possible 
application to more complex systems, the computational resources
required by the quantum tomographic approach are high in general.  
%
%

On the side of the exact techniques, the Chebyshev approach largely 
outperforms the standard Crank-Nicholson scheme, both in computation speed 
(usable time step) and treatable system sizes (only matrix-vector 
multiplications were required).
We conclude that the Chebyshev approach represents a
very efficient and reliable tool to determine the quantum dynamics 
even for rather complex interacting many-particle systems. 
%



\end{document}